# Spin Dynamics and Magnetoelectric Coupling Mechanism of $Co_4Nb_2O_9$

Guochu Deng[a,1], Yiming Cao[b,c], Wei Ren[b], Shixun Cao[b,1], Andrew J. Studer[a], Nicolas Gauthier[d], Michel Kenzelmann[d], Gene Davidson[a], Kirrily C. Rule[a], Jason S. Gardner[e], Paolo Imperia[a], Clemens Ulrich[f], Garry J. McIntyre[a]

[a] *Australian Nuclear Science and Technology Organisation, New Illawarra Road, Lucas Heights NSW 2234, Australia*

[b] *Department of Physics, International Centre of Quantum and Molecular Structures, and Materials Genome Institute, Shanghai University, Shanghai 200444, China*

[c] *Centre for Magnetic Materials and Devices & Key Laboratory for Advanced Functional and Low Dimensional Materials of Yunnan Higher Education Institute, Qujing Normal University, 655011, China*

[d] *Laboratory for Scientific Developments & Novel Materials, Paul Scherrer Institut, CH-5232 Villigen-PSI, Switzerland.*

[e] *Neutron Group, National Synchrotron Radiation Research Centre, Hsinchu 30076, Taiwan*

[f] *School of Physics, The University of New South Wales, Kensington NSW 2052, Australia.*

**Abstract**

Neutron powder diffraction experiments reveal that $Co_4Nb_2O_9$ forms a noncollinear in-plane magnetic structure with $Co^{2+}$ moments lying in the *ab* plane. The spin-wave excitations of this magnet were measured by using inelastic neutron scattering and soundly simulated by a dynamic model involving nearest and next-nearest neighbour exchange interactions, in-plane anisotropy and the Dzyaloshinskii-Moriya interaction. The in-plane magnetic structure of $Co_4Nb_2O_9$ is attributed to the large in-plane anisotropy while the noncollinearity of the spin configuration is attributed to the Dzyaloshinskii-Moriya interaction. The high magnetoelectric coupling effect of $Co_4Nb_2O_9$ in fields can be explained by its special in-plane magnetic structure.

---

[1] Corresponding author, Email: guochu.deng@ansto.gov.au; sxcao@shu.edu.cn

**Introduction**

Multiferroic materials have been extensively investigated over the last two decades due to their potential applications in future electronics such as storage devices, sensors, etc.[1] In spite of great progress in the application-oriented research in this field,[2,3] fundamental studies on the mechanism of magnetoelectric (ME) coupling and the origin of multiferroicity of various multiferroic systems are full of challenges due to the inherent difficulties of this topic.[4-8]

Recently, $Co_4Nb_2O_9$ (CNO) was reported to demonstrate a high ME coupling coefficient.[9,10] This compound belongs to the space group $P\bar{3}c1$. As shown in Fig. 1 (a), $Co^{2+}$ ions split into Co1 and Co2 sites and form chains along the $c$ axis with alternating spacings. The octahedra on the Co1 sites connect into a nearly planar network by edge-sharing while those on the Co2 sites join into a buckled network by corner-sharing. These two networks stack along the $c$ axis alternatively. $Co_4Nb_2O_9$ undergoes an antiferromagnetic phase transition at a low temperature $T_N$ (~28K). Surprisingly, the observed ME coupling effect[9,10] and magnetization measurements[11] are contradictive to the magnetic structures previously proposed by Bertaut *et al.*[12] and recently by Khanh *et al.*[13] The former claimed a collinear magnetic structure with $Co^{2+}$ moments aligning along the $c$ axis[12] while the latter suggested that magnetic moments lie primarily in the $ab$ plane with certain canting angles towards the $c$ axis.[13] We have measured the magnetization of single-crystal samples along different crystallographic axes. The results clearly shows a cusp in the data measured along the $a$ axis but no abnormality along the $c$ axis near $T_N$.[11,14] This suggests that the magnetic moments *do not* have components along the $c$ axis, which is incompatible with the two magnetic structure models mentioned above. Thus, the previous interpretations to the ME coupling mechanism in CNO should be reconsidered too due to the lack of accurate magnetic structure for this compound.

In this paper, we discover an in-plane noncollinear magnetic structure for CNO by using neutron powder diffraction and irreducible representation analysis. We also propose a dynamic model, which simulates the inelastic neutron scattering data. It indicates that large in-plane anisotropy primarily causes the easy-plane magnetic structure, the Goldstone mode, and the gapped mode in the spin-wave spectrum. Similarly, the Dzyaloshinskii-Moriya (DM) interaction is crucial to the noncollinearity of the spin configuration. Bearing this in-plane noncollinear magnetic structure in mind, we discuss the origin of the ME coupling of CNO in magnetic fields in depth.

**Experiment**

$Co_4Nb_2O_9$ powder and single-crystal samples were prepared by a solid-state reaction and the optical floating-zone method, respectively, at Shanghai University.[11] The neutron powder diffraction experiment was carried out on the high-flux neutron diffractometer WOMBAT[15] at the OPAL reactor of the Australian Nuclear Science and Technology Organisation (ANSTO) using a nominal wavelength of λ=2.41 Å in the temperature range from 5K to 50K. The irreducible representation analysis was performed using SARAh.[16]. The crystal and magnetic structure models were refined by the Rietveld method using the program Fullprof.[17, 18]

Inelastic neutron scattering experiments were performed on both the thermal-neutron and cold triple-axis spectrometers, TAIPAN[19] and SIKA[20] at OPAL, respectively. On TAIPAN, the instrument was configured with an open-open-open-open collimation and a double focus monochromator and analyser using the fixed $E_f$ = 14.87 meV. On SIKA, 60'-60'-60'-60' collimation was used with $E_f$ = 8.07 meV. A single-crystal sample of CNO with a mass of about 2 g was mounted on an Al sample holder and aligned in the *ac* plane for the experiments. Most of the data were collected near the antiferromagnetic zone centre (100) by constant-Q scans. The spin-wave dispersion was simulated using the spin wave calculation package SpinW.[21]

**Magnetic order**

Subtracting the neutron powder diffraction pattern measured above $T_N$ from the one below $T_N$ clearly shows magnetic Bragg peaks in the low-Q range. All magnetic peaks overlap with the nuclear Bragg peaks, indicating that the propagation vector of the ordered magnetic phase is ***k*** = (0, 0, 0). The magnetic $Co^{2+}$ ions occupy two different sites, namely, (1/3, 2/3, 0.017) and (2/3, 1/3, 0.3078), in CNO. The detailed crystal structure is given in the TABLE II of the Supplemental Material.

According to the analysis, the magnetic representation of CNO can be decomposed into six irreducible representations (in short, Irreps) $\Gamma = \Gamma_1 + \Gamma_2 + \Gamma_3 + \Gamma_4 + 2\Gamma_5 + 2\Gamma_6$, as shown in TABLE I of the Supplemental Material.[22] Among them, the basis vectors of the first four Irreps ($\Gamma_1$ to $\Gamma_4$) are parallel or antiparallel to the hexagonal axis since these Irreps still keep the three-fold rotation symmetry. These four Irreps denote respectively four spin configurations ( (++--), (+--+), (++++), and (+-+-)) on one chain. The other two Irreps ($\Gamma_5$ and $\Gamma_6$) have the basis vectors in the $ab$ plane with breaking of the three-fold rotation symmetry. In other words, magnetic moments of these two Irreps can lie only in the $ab$ plane.

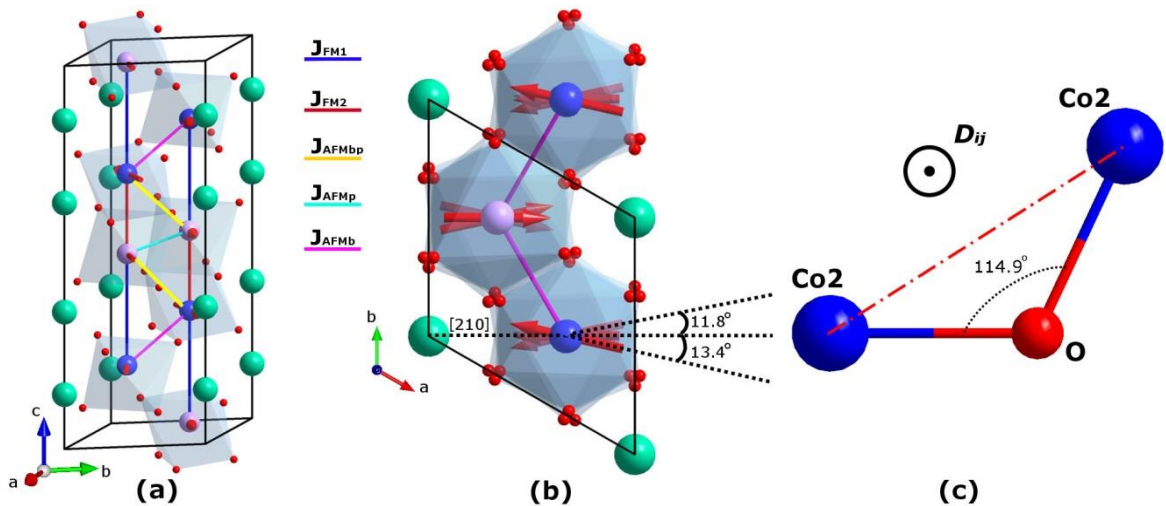

Fig. 1 (a) Crystal and magnetic structure models of CNO; (b) The canting of magnetic moments in the $ab$ plane; (c) the bond angle of Co2-O1-Co2, which is the origin of DM interaction; In these figures, red spheres denote oxygen atoms, green ones denote niobium atoms, and purple and blue represent Co1 and Co2 atoms, respectively. The red arrows show the orientation of magnetic moments, which are in the $ab$ plane with certain canting angles. The colour lines between $Co^{2+}$ shows the exchange interactions. The values in (b) are the angles between the magnetic moments of Co2 and the [210] direction.

Considering all these Irreps suggested by SARAh[16], we found that $\Gamma_6$ gives the best fit to the observed diffraction pattern (Fig. 2). This refinement generates a noncollinear magnetic structure as shown in Fig. 1 (a) and (b). The agreement indices of the refinement are $R_p$ = 3.23, $R_{wp}$ = 4.11, magnetic $R_B$= 7.08 and $\chi^2$=2.92. The magnetic moments on both Co1 and Co2 sites lie in the *ab* plane. However, the magnetic moments within each set of sites (Co1 and Co2) are not collinear in the plane, but slightly cant from each other. The average direction of each set of moments is roughly along the [210] direction. The canting angle of the neighbour Co1 moments is 1.3±0.1° while the canting angle of the Co2 moments is much larger, 25.2(1)° (=11.8°+13.4°, see Fig. 1(b)). The refined magnetic moments on the Co1 and Co2 sites are 2.32(1) $\mu_B$ and 2.52(1) $\mu_B$, respectively.

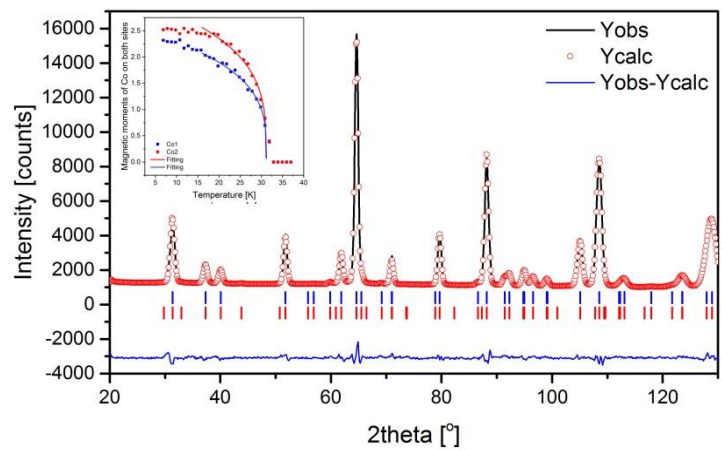

Fig. 2 Neutron powder diffraction pattern at 5K from the experiment (black curve) and the Rietveld refinement (red circles) with the Irreps $\Gamma_6$; The red tick marks below the data show the nuclear Bragg peak positions of CNO while the blue tick marks show the magnetic Bragg peak positions of the antiferromagnetic phase. Inset: The temperature dependencies of the Co1 and Co2 magnetic moments obtained by fitting the diffraction data. The critical exponents are 0.29(2) and 0.28(2) for the Co1 and Co2 sites, respectively.

Our magnetic structure model is different from the two previously reported magnetic structures[12, 13], although it is slightly similar to the one proposed by Khanh *et al.*[13] It is worthwhile to compare Khanh *et al.*'s model and ours in detail. First of all, the main magnetic components of both models are not along the *c* axis. According to our results, the magnetic moments are purely in the *ab* plane while Khanh *et al.*'s are nearly in the *ab* plane with a canting towards the *c* axis. Secondly, the in-plane magnetic components of both models are roughly aligned along the [210] direction. In our model, however, the magnetic moments cant in the *ab* plane. In contrast, the moments in Khanh *et al.*'s model are collinear in the *ab* plane.[13] Thirdly, the moment magnitudes are similar on both the Co1 and Co2 sites in our model, while the magnetic moment on the Co2 sites (3.5 $\mu_B$) is far larger than the one on the Co1 sites (2.6 $\mu_B$) in Khanh *et al.*'s model. Finally, the magnetic moments canting towards the *c* axis in Khanh *et al.*'s model[13] is completely contradictive to our magnetization measurements and incompatible with the irreducible representation analysis (see TABLE I in the Supplemental Material[22]). A recent theoretical calculation by Solovyev *et al.*[23] indicates that the canting angle should be no more than 2º if canting exists, which is substantially smaller than the canting angles 22º and 24º in Khanh *et al.*'s model. This theoretical work hints that the in-plane magnetic structure is more favourable with regard to energy. Refining our data using Khanh *et al.*'s model gave $R_p$ = 3.56, $R_{wp}$ = 4.86, magnetic $R_B$ = 11.52 and $\chi^2$ = 4.07, which are considerably larger than the corresponding values obtained with our model. Taking all the arguments above into account, we conclude that the magnetic structure of CNO should have no canting moment along the *c* axis.

**Spin dynamics**

The results from the inelastic neutron scattering experiment of CNO are presented in Fig. 3 (a) and (b). In Fig. 3(a), the dispersion curves of the spin-wave excitation of CNO were measured along the $Q_H$ direction at ~ 5 K. The experimental data show two branches along both $Q_H$ and $Q_L$ directions after convolution fitting with the instrumental resolution. The lower branch corresponds to a Goldstone mode, which goes to zero at the zone centre and about 3.5meV at the zone boundary. The upper branch is a gapped mode, which has an energy gap of about 3.1meV at the zone centre. Along the $Q_L$ direction, both the gapped mode and the Goldstone mode are visible along the $Q_L$ direction, as shown in Fig. 3(b). The Goldstone mode extends to slightly higher energy (~4.3 meV) at the zone boundary along the $Q_L$ direction. The intensity of this mode substantially drops in the range $Q_L$ > 1.0 r.l.u. while the gapped mode still propagates in the same range with gradually decreasing intensity. Additionally, some optical modes are observed at a higher-energy range from 6 to 8 meV along the $Q_L$ direction. Fig. 3(c) shows the energy scans at three different Q positions with higher resolution on SIKA, where both the Goldstone and gapped modes can be clearly seen.

The rather large energy gap ( > 3 meV) indicates a contribution not from a DM interaction but from single-ion anisoropy since the DM interaction is usually much weaker. The coexistence of the Goldstone and gapped modes in CNO excludes the possibility that CNO is an easy-axis magnet like MnF$_2$.[24] The fact that the magnetic moments of CNO lie in the *ab* plane indicates it is an easy-plane magnet. Clearly, the Goldstone mode corresponds to an in-plane spin excitation,[25] and the gapped mode is related to an out-of-plane spin excitation.

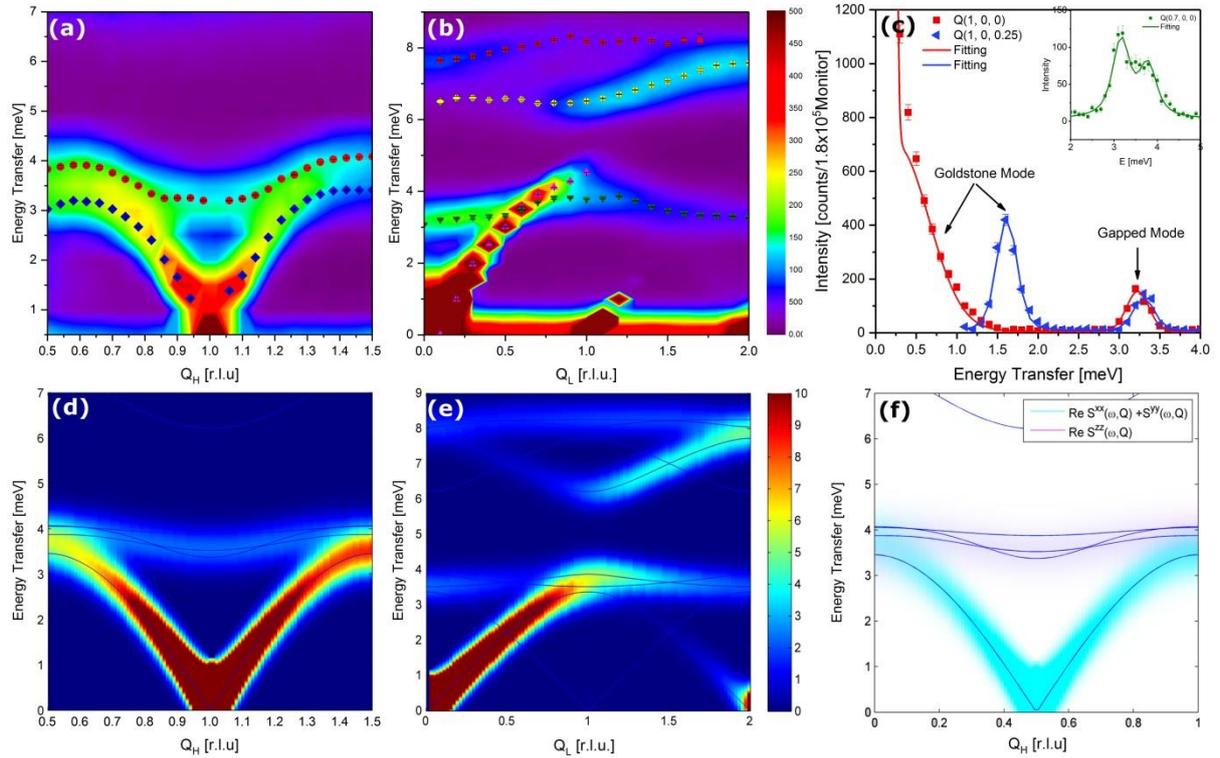

Fig.3 Spin-wave dispersion measured near the zone centre (100) along the (a) $Q_H$ and (b) $Q_L$ directions in CNO single crystal (from TAIPAN). The dots are the fitted peak positions by convoluting with the instrument resolution. (c) The energy scans at different longitudinal or transverse Q positions (from SIKA). The solid lines correspond to the fittings where the convolution with the instrument resolution was taken into account. (d) and (e) shows the simulated spin-wave spectra along the $Q_H$ and $Q_L$ directions, respectively, using the dynamic model described in the text. (f) The simulated in-plane and out-of-plane components of the excitation.

In order to build a dynamic model to understand the dispersion data in Fig. 3 (a) and (b), we carefully consider the interactions between nearest and next-nearest neighbours (NN and NNN, respectively) of the $Co^{2+}$ ions in CNO. Firstly, our diffraction data indicate the distances between NN and NNN Co1 and Co2 ions are 2.92 Å and 4.15 Å along the *c* axis, respectively. The superexchange pathways for the NN and NNN interactions ($J_{FM1}$ and $J_{FM2}$) run through the bonds Co1-O-Co2 and Co1-O-O-Co2, respectively (see Fig. S1 in the Supplemental Material[22]). All bond angles on both these pathways are close to 90° (see TABLE II in the Supplemental Material[22]). According to the Goodenough-Kanamori-Anderson (GKA) rule, both $J_{FM1}$ and $J_{FM2}$ should therefore be ferromagnetic.[26] This is consistent with our magnetic structure, in which magnetic moments on the same chain align roughly in parallel.

Secondly, we consider the exchange pathways on the planar buckled networks, and in between. The interaction ($J_{FMp}$) between any two neighbour Co1 ions in the planar network takes place through a similar ~90° Co1-O-Co1 superexchange pathway, which should be ferromagnetic according to the GKA rule.[26] However, the magnetic structure shows that the magnetic moments on these two sites are roughly antiparallel. Thus, we speculate that this interaction might be very weak. The NN interaction ($J_{AFMb}$) on the buckled network goes through a Co2-O-Co2 pathway with a bond angle of ~ 115° (see Fig. 1(c)), resulting in an antiferromagnetic interaction according to the GKA rule.[26] The exchange $J_{AFMbp}$ between Co1 and Co2 takes place through a Co1-O-Co2 bond with an angle of ~ 129°, supposing to be antiferromagnetic. The coloured lines in Fig. 1 (a) and (b) show all the exchange pathways discussed above.[22]

Considering the single-ion anisotropy and DM interaction as well, the Hamiltonian for this Heisenberg magnet reads:

$$H = J_{FM1} \sum_{<i,j>} \vec{S}_i^I \cdot \vec{S}_j^{II} + J_{FM2} \sum_{<i,j>} \vec{S}_i^I \cdot \vec{S}_j^{II} + J_{FMp} \sum_{<i,j>} \vec{S}_i^I \cdot \vec{S}_j^I + J_{AFMb} \sum_{<i,j>} \vec{S}_i^{II} \cdot \vec{S}_j^{II}$$
$$+ J_{AFMbp} \sum_{<i,j>} \vec{S}_i^I \cdot \vec{S}_j^{II} + \sum_{Co1} D_1 (S_{iZ}^I)^2 + \sum_{Co2} D_2 (S_{iZ}^{II})^2 + \sum_{<i,j>} D_{ij} \cdot (\vec{S}_i^{II} \times \vec{S}_j^{II})$$

where $J_{FM1}$ and $J_{FM2}$ are the ferromagnetic exchange interactions of NN and NNN along $c$; $J_{FMp}$, $J_{AFMb}$ and $J_{AFMbp}$ are the antiferromagnetic exchange interactions of the 'planar' networks, the 'buckled' networks and between these two networks, respectively; I and II denote the Co1 and Co2 sites, respectively; $D_1$ and $D_2$ are positive, indicating easy-plane anisotropies of the Co1 and Co2 sites, respectively; and $D_{ij}$ is the DM interaction.

Since Co1 and Co2 moments are quite similar in size and orientation, we assume that $D_1$ and $D_2$ take the same value. The DM interaction is temporarily neglected because it is normally much weaker than the exchange parameters and anisotropy. With this simplified model, we simulated the spin-wave dispersion by using SpinW.[21] The optimized simulation results along $Q_H$ and $Q_L$ directions are shown in Fig. 3 (d) and (e), respectively. The parameters used for this simulation are $J_{FM1}$ = -0.70meV, $J_{FM2}$ = -0.15meV, $J_{FMp}$ = -0.16meV, $J_{AFMb}$ = 0.42meV, $J_{AFMbp}$ = 0.52meV and $D_1 = D_2$ = 1.8meV. These results indicate that the single-ion anisotropy is quite large in CNO, much larger than other exchange interactions. It is not unusual to observe large spin anisotropy in cobaltites. For example, $Ba_2CoGe_2O_7$ undergoes an easy-plane antiferromagnetic phase transition at ~ 7 K.[27] According to the inelastic neutron scattering experiment and theoretical simulation,[27, 28] the NN interaction $J_{AFM}$ in $Ba_2CoGe_2O_7$ is about 0.2 meV while the single-ion anisotropy $D$ is about 1.15 meV. The large anisotropy value is comparable to that of CNO in this study.

Comparing with Fig. 3 (a) and (b), the simulations in Fig.3 (d) and (e) reproduce almost all the features of the experimental results, for example, the Goldstone and gapped modes. In Fig. 3(f), the Goldstone mode corresponds to the in-plane spin excitation while the gapped mode originate from the out-of-plane spin excitation.[22] The intensities the simulated dispersion curves also agree well with the experiment over the whole Q range, especially, reproducing the low intensity in the range $Q_L$ > 1.0 r.l.u. This consistency suggests that our model describes this magnetic system very well.

In this model, a collinear magnetic ground state is established by using the simulated annealing method in SpinW. This disagrees with our magnetic structure model from powder diffraction. If we introduce a small DM interaction into this model through the Co2-O-Co2 path (see Fig. 1(c) and the further discussion in next section), the simulated annealing produces a magnetic structure with different canting angles for moments on both the Co1 and Co2 sites.[21] The canting angles are dependent on the value of the DM interaction. A value of DM ≈ 0.22 meV is able to generate similar canting angles for the Co1 and Co2 sites (roughly ~5° and ~23°, respectively), close to our magnetic structure model. The comparison between our magnetic structure model and the simulated ground-state magnetic structure is shown in Fig. S2 in the Supplemental Material.[22] The similarity between them strongly indicates that the DM interaction is indispensable to understand the noncollinearity of the magnetic structure of CNO. It is worthwhile to point out that introducing the DM interaction as above does not visibly change the simulated spin-wave dispersion in Fig. 3 (d), (e) and (f).

**Origin of magnetoelectric effect**

In our magnetic structure model, $S_i$ and $S_j$ are in the *ab* plane and noncollinear. Thus, their product should be a vector along the *c* axis. Since $e_{ij}$ is a unit vector from $S_i$ to $S_j$, which is in the (110) plane (or any equivalent planes), the polarization *P* should be along [110] (or any equivalent) direction because *P* is proportional to $e_{ij} \times (S_i \times S_j)$ according to the spin-current model.[7] However, *P* is cancelled by the neighbour polarization *P'* due to the antiparallel spin configuration. This agrees with the zero polarization at zero field reported in the literature.[10,13]

What happens when applying a magnetic field? According to Chubokov,[29] the magnetic moments of a hexagonal in-plane magnet stay in-plane even when applying a strong out-of-plane magnetic field if its in-plane anisotropy *D* is much larger than the dominant exchange interaction. Therefore, we only consider the in-plane field effect here. In an in-plane magnetic field, magnetic moments will slightly rotate to reduce their angles to the external field direction, which causes non-equivalent *P* and *P'*. Consequently, the total polarization will be nonzero. In a higher field, both *P* and *P'* point to the same direction due to the spin-flip transition, resulting in an enhanced polarization. More details on the field effect are given in the Supplemental Material (see Fig. S3).[22] *According to this understanding, interestingly, no matter along which direction the external field is, the total polarization always remains along the [110] or any equivalent directions.* This is determined by the magnetic structure of CNO itself. Our analysis is strongly supported by the fact that the maximal polarization is always observed along the [110] direction, as reported by Khanh *et al.*[13]

The spin-current model is also called the inverse-DM effect. It is interesting to know which bond causes this effect in CNO. The pathway Co2-O-Co2 forms a triangle with a bond angle of ~115°. Co2 moments have larger canting angles than Co1 moments on the Co1-O-Co1 bond. Therefore, we conclude that the DM interaction takes place through the Co2-O-Co2 bond rather than through the the Co1-O-Co1 bond. The dynamic model involving DM on the Co2-O-Co2 bond generates a noncollinear magnetic ground state, in agreement with our magnetic structure model determined from powder diffraction (see Fig. S2 in the Supplemental Material[22]). The spin canting angles on the Co1 sites are smaller than those on the Co2 sites. On the contrary, a DM interaction on Co1-O-Co1 bond produces an opposite result. Therefore, it is the Co2-O-Co2 bond that contributes to the DM interaction in CNO, which finally causes the strong ME coupling.

**Conclusion**

The magnetic structure and spin dynamics of CNO were studied by neutron powder diffraction and inelastic neutron scattering, respectively. The dynamic behaviours such as the existence of the Goldstone and gapped modes reveal that CNO is an easy-plane magnet with large easy-plane anisotropy. The DM interaction causes the noncollinear in-plane magnetic structure of CNO. The high ME coupling effect of CNO can be explained entirely by the spin-current model. Our proposed mechanism can be widely used to explain the ME coupling effect in other easy-plane magnets. Therefore, this study also provides a guideline to search for novel ME material candidates in hexagonal easy-plane magnets.

**Acknowledgement**

We thank ANSTO for the allocation of neutron beam time on WOMBAT, SIKA and TAIPAN (P3843, P3844, P4348, and P4851). This work is supported by the National Natural Science Foundation of China (NSFC, Nos. 11774217, 51372149, 51672171), the National Key Basic Research Program of China (Grant No. 2015CB921600), Eastern Scholar Program from Shanghai Municipal Education Commission, and the research grant (No.16DZ2260600) from Science and Technology Commission of Shanghai Municipality.

# Supplemental Material

In this Supplemental Material, we show the irreducible-representation analysis, the magnetic structure refinement parameters, the superexchange pathways, the Dzyaloshinskii-Moriya (DM) effect, and the magnetoelectric (ME) effect under magnetic fields for $Co_4Nb_2O_9$ (CNO). Section A shows the basis vectors of the irreducible representations for the magnetic ions $Co^{2+}$ in CNO and the refinement parameters. In Section B, the superexchange pathways are displayed and analysed under the Goodenough-Kanamori-Anderson (GKA) rule.[1] In Section C, the impact of the DM interaction on the magnetic ground state of CNO is explained. Section D presents the mechanism of the ME coupling effect under magnetic fields.

## A. Irreducible representation analysis and magnetic structure refinement

According to the software SARAh,[2] the magnetic representation of magnetic ions in CNO can be reduced into six irreducible representations (Irreps). The basis vectors of these Irreps are listed in TABLE I. From these basis vectors, it is clearly seen that the Irreps #1 ~ #4 have the magnetic moments aligning along the $c$ axis while the Irreps #5 and #6 have the magnetic moments in the $ab$ plane. Each of them has four basis vectors. The refined parameters for the crystal structure and magnetic structure of CNO are shown in TABLE II.

TABLE I. Basis vectors of all irreducible representations and the corresponding spin configuration

| IR | Co1 | Co2 | Co3 | Co4 | Configuration |
|---|---|---|---|---|---|
| #1 | (0  0  3) | (0  0  -3) | (0  0  3) | (0  0  -3) | AFM along $c$ |
| #2 | (0  0  3) | (0  0  -3) | (0  0  -3) | (0  0  3) | AFM along $c$ |
| #3 | (0  0  3) | (0  0  3) | (0  0  3) | (0  0  3) | FM along $c$ |
| #4 | (0  0  3) | (0  0  3) | (0  0  -3) | (0  0  -3) | AFM along $c$ |
| #5 | (1.5, 0, 0) | (-1.5, -1.5, 0) | (1.5, 0, 0) | (-1.5, -1.5, 0) | In $ab$ plane |
| #5 | (0, 1.5, 0) | (0, 1.5, 0) | (0, 1.5, 0) | (0, 1.5, 0) | In $ab$ plane |
| #5 | (0.866, 1.732, 0) | (-0.866, 0.866, 0) | (0.866, 1.732, 0) | (-0.866, 0.866, 0) | In $ab$ plane |
| #5 | (-1.732, -0.866, 0) | (1.732, 0.866, 0) | (-1.732, -0.866, 0) | (1.732, 0.866, 0) | In $ab$ plane |
| #6 | (1.5, 0, 0) | (-1.5, -1.5, 0) | (-1.5, 0, 0) | (1.5, 1.5, 0) | In $ab$ plane |
| #6 | (0, 1.5, 0) | (0, 1.5, 0) | (0, -1.5, 0) | (0, -1.5, 0) | In $ab$ plane |
| #6 | (0.866, 1.732, 0) | (-0.866, 0.866, 0) | (-0.866, -1.732, 0) | (0.866, -0.866, 0) | In $ab$ plane |
| #6 | (-1.732, -0.866, 0) | (1.732, 0.866, 0) | (1.732, 0.866, 0) | (-1.732, -0.866, 0) | In $ab$ plane |

TABLE II. Refinement results of the crystal structure and magnetic structure of CNO. The neutron diffraction data were taken at a temperature of 5K using a wavelength of λ = 2.41 Å

| Space group | a | b | c | α | β | γ |
|---|---|---|---|---|---|---|
| P$\bar{3}$c1 | 5.14711(2) | 5.14711(2) | 14.0724(3) | 90 | 90 | 120 |
| **Atom** | *x* | *y* | *z* | Biso | | |
| Nb1 | 0 | 0 | 0.3577(1) | 0.5696(5) | | |
| Co1 | 1/3 | 2/3 | 0.0170(2) | 0.4240(3) | | |
| Co2 | 1/3 | 2/3 | 0.3078(2) | 0.4240(3) | | |
| O1 | 0.2901(1) | 0 | 0.25 | 0.5011(3) | | |
| O2 | 0.3437(2) | 0.3205(3) | 0.0857(3) | 0.7463(5) | | |
| **Magnetic atoms** | *x* | *y* | *z* | Magnetic Moment ($\mu_B$) | | | |
| | | | | $M_a$ | $M_b$ | $M_c$ | Total |
| Co1 | 1/3 | 2/3 | 0.0170(2) | 2.677(1) | 1.312(1) | 0 | 2.319(5) |
| Co1 | 2/3 | 1/3 | 0.4830(2) | -2.677(1) | -1.365(1) | 0 | 2.319(5) |
| Co1 | 2/3 | 1/3 | 0.9830(2) | -2.677(1) | -1.312(1) | 0 | 2.319(5) |
| Co1 | 1/3 | 2/3 | 0.5170(2) | 2.677(1) | 1.365(1) | 0 | 2.319(5) |
| Co2 | 1/3 | 2/3 | 0.3078(2) | 2.842(1) | 1.953(1) | 0 | 2.519(5) |
| Co2 | 2/3 | 1/3 | 0.1922(2) | -2.842(1) | -0.889(1) | 0 | 2.519(5) |
| Co2 | 2/3 | 1/3 | 0.6922(2) | -2.842(1) | -1.953(1) | 0 | 2.519(5) |
| Co2 | 1/3 | 2/3 | 0.8078(2) | 2.842(1) | 0.889(1) | 0 | 2.519(5) |

**B. The superexchange pathways of $Co^{2+}$**

The interactions between the NN and NNN $Co^{2+}$ ions go through the superexchange pathways. In Fig. S1, these superexchange pathways are indicated along the *c* axis, in the planar network and the buckled network in the direction perpendicular to the *c* axis. From the crystal structure refinement of CNO, the bond angles on these pathways can be determined (See TABLE III). According to the GKA rule,[1] it is possible to determine of which type the exchange interactions are, namely, ferromagnetic or antiferromagnetic.

TABLE III The NN and NNN superexchange pathways of $Co^{2+}$

| Pathway | Exchange | Direction | Angle | Interaction |
|---|---|---|---|---|
| Co1-O-Co2 | $J_{FM1}$ | Along *c* | 84.14(1)° | FM |
| Co1-O-O-Co2 | $J_{FM2}$ | Along *c* | 89.38(1)°, 93.63(1)° | FM |
| Co1-O-Co1 | $J_{FMp}$ | Planar network | 90.68(1)° | Weak FM |
| Co2-O-Co2 | $J_{AFMb}$ | Buckled network | 114.95(1)° | AFM |
| Co1-O-Co2 | $J_{AFMbp}$ | Planar-Buckled network | 129.68(1)° | AFM |

For example, the NN and NNN neighbors along the $c$ axis have a bond angle of about 90° along the superexchange pathway. Thus, both of them should be ferromagnetic. This is consistent with the magnetic structure obtained by neutron powder diffraction. Furthermore, the Co1-O-Co1 pathway on the planar network also has a ~90° bond angle (see Fig. S1(b)). The Co moments are expected to be ferromagnetic as well. However, the magnetic moments on these two Co1 sites are aligned in an anti-parallel way according to the magnetic structure. This indicates that this interaction is relatively weak. In contrast, the Co2-O-Co2 pathway on the buckled network should be antiferromagnetic according to the GKA rule because the bond angle is ~ 115°. Additionally, the interaction on the Co1-O-Co2 bond between the buckled and planar networks has a bond angle of ~ 128°, which is supposed to be antiferromagnetic.

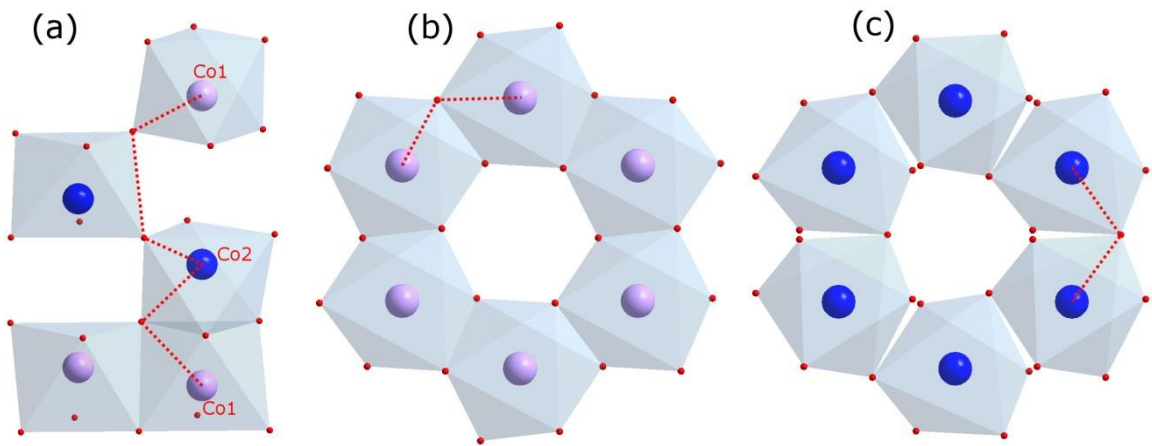

Fig. S1 (a) The superexchange pathways of the NN and NNN neighbours along the $c$ direction; (b) The superexchange pathway on the nearly planar network of $(Co1)O_6$ octahedra; (c) The superexchange pathway on the buckled network of $(Co2)O_6$ octahedra. The red lines show the superexchange pathways and the bond angles.

## C. The DM interaction effect on the magnetic ground state

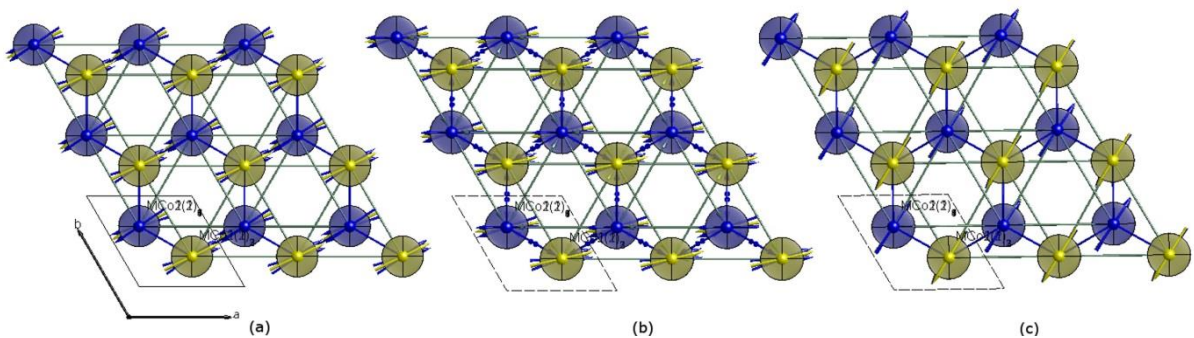

Fig. S2 (a) The magnetic structure determined by neutron powder diffraction; (b) The ground-state magnetic structure by the simulated annealing method on the basis of the proposed dynamic model with DM interaction; (c) The ground-state magnetic structure by the simulated annealing method on the basis of the proposed spin-dynamics model without DM interaction;

The dynamic model proposed in the article is able to explain the observed spin-wave excitation. However, it is not able to extract straightforwardly the DM interaction because the DM interaction is normally much weaker and has less obvious impact on the feature of the spin-wave dispersion curves. However, if we simulate the ground-state magnetic structure of CNO using the dynamic model discussed in the article without involving the DM interaction, the simulated annealing procedure finally produces a collinear magnetic structure, as shown in Fig. S2 (c). If we add a DM interaction on the Co2-O-Co2 pathway shown in Fig. S1 (c) to the dynamic model, the simulated ground-state magnetic structure becomes a noncollinear one. The in-plane canting angle strongly depends on the DM interaction value. Fig. S2 (b) shows the simulated ground-state magnetic structure of CNO with optimized the DM interaction. This result looks very similar to the neutron-powder diffraction result in Fig. S2 (a). This demonstrates that the DM interaction is indispensable to the dynamic model in order to explain the noncollinear magnetic structure of CNO.

### D. The mechanism of the ME coupling in CNO

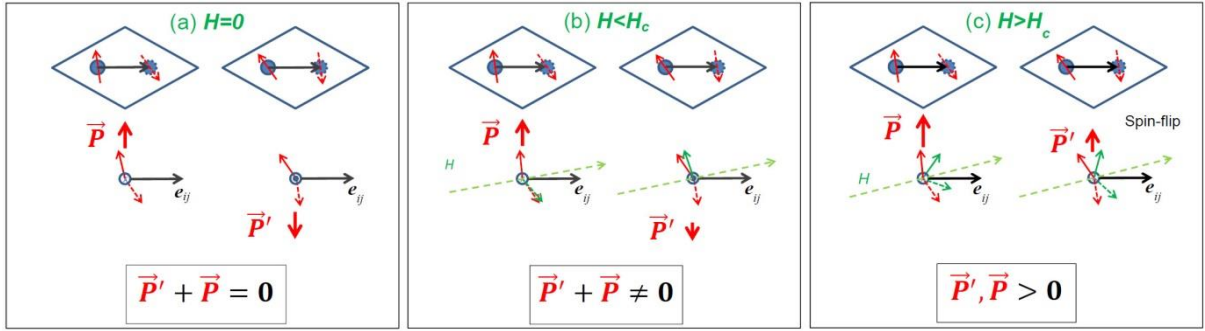

Fig. S3 (a) The local and total polarizations at zero field; (b) The local and total polarizations at a small in-plane magnetic field lower than the critical field $H_c$; (c) The local and total polarizations at a small in-plane magnetic field higher than the critical field $H_c$.

As been discussed above, CNO has the in-plane anisotropy $D$ much larger than the exchange interactions $J$. According to the previous theoretical work by Chubokov[3], the in-plane magnetic structure will be quite stable in a magnetic field along the $c$ axis if $D \gg J$. Therefore, we focus our discussion on the in-plane magnetic field effect here. Fig. S3 presents the scenario for the ME coupling effect of CNO under different in-plane magnetic fields. In the zero field (see Fig. S3(a)), the two local polarizations are just antiparallel and cancel out each other, resulting in a zero total polarization, which is consistent with the reported results.[4] When applying a small in-plane magnetic field, the spins start to slightly rotate to the external field direction in the easy plane. This rotation will have non-equivalent effects on the two local polarizations. Thus, the total polarization becomes nonzero. When increasing the magnetic field above the critical field $H_c$, a spin-flip transition takes place. All the magnetic moment will have a component along the field direction. In this case, the two polarizations (*P* and *P'*) have the same direction, and the total polarization will increase. The most important feature of this model is that the polarization is always along the [110] direction (or any equivalent directions) no matter in which direction the external magnetic field is applied. This is strongly supported by the fact that the maximum polarization is always observed along the [110] or equivalent direction[5].